%% file: preprint.tex
\documentstyle[12pt,aaspp4]{article}

\input psfig2

\def\deg{\ifmmode ^{\circ}
         \else $^{\circ}$\fi}
\def\pdeg{\ifmmode $\setbox0=\hbox{$^{\circ}$}\rlap{\hskip.11\wd0 .}$^{\circ}
          \else \setbox0=\hbox{$^{\circ}$}\rlap{\hskip.11\wd0 .}$^{\circ}$\fi}
\def\arcs{\ifmmode {^{\scriptscriptstyle\prime\prime}}
          \else $^{\scriptscriptstyle\prime\prime}$\fi}
\def\arcm{\ifmmode {^{\scriptscriptstyle\prime}}
          \else $^{\scriptscriptstyle\prime}$\fi}
\newdimen\sa  \newdimen\sb
\def\parcs{\sa=.07em \sb=.03em
     \ifmmode $\rlap{.}$^{\scriptscriptstyle\prime\kern -\sb\prime}$\kern -\sa$
     \else \rlap{.}$^{\scriptscriptstyle\prime\kern -\sb\prime}$\kern -\sa\fi}
\def\parcm{\sa=.08em \sb=.03em
     \ifmmode $\rlap{.}\kern\sa$^{\scriptscriptstyle\prime}$\kern-\sb$
     \else \rlap{.}\kern\sa$^{\scriptscriptstyle\prime}$\kern-\sb\fi}
\def\gtorder{\mathrel{\raise.3ex\hbox{$>$}\mkern-14mu
             \lower0.6ex\hbox{$\sim$}}}
\def\ltorder{\mathrel{\raise.3ex\hbox{$<$}\mkern-14mu
             \lower0.6ex\hbox{$\sim$}}}
\newcommand{\kms}{\mbox{ km~s$^{-1}$}}

\begin{document}

\title{SBS 0909+532: A New Double Gravitational Lens or Binary Quasar?
\footnote{Observations reported here were obtained, in part, at 
MDM Observatory,
a consortium of the University of Michigan, Dartmouth College
and the Massachusetts Institute of Technology.}
\footnote{This research made use of the NASA/IPAC Extragalactic Database (NED)
  which is operated by the Jet Propulsion Laboratory, Caltech, under contract
  with the National Aeronautics and Space Administration}
\footnote{We have made use in part of finder chart(s)
         obtained using the Guide Stars Selection System Astrometric Support
         Program developed at the Space Telescope Science Institute (STScI is
         operated by the Association of Universities for Research in Astronomy,
         Inc., for NASA) }
}
\author{Christopher S. Kochanek}
\author{Emilio E. Falco}
\author{Rudolf Schild}
\author{Adam Dobrzycki}
\affil{Harvard-Smithsonian Center for Astrophysics\protect \\ 
       60 Garden Street \protect \\
       Cambridge MA 02138 }
\author{Dieter Engels}
\author{Hans-J\"urgen Hagen}
\affil{ Hamburger Sternwarte \protect \\
        Gojenbergsweg 112, D-21029 Hamburg \protect \\
        Germany }
\authoremail{ckochanek@cfa.harvard.edu}
\authoremail{falco@cfa.harvard.edu}

\begin{abstract}
The $z=1.377$, $B=17.0$ mag quasar SBS 0909+532 A, B is a double 
with two images separated by $\Delta \theta= 1\parcs107\pm0\parcs006$.  
Because the faint image has an emission line at the same 
wavelength as the MgII 2798 \AA\ emission line of the quasar, 
and lacks the broad MgIb absoption feature expected for a star
with the same colors (a K star), we conclude that image B is 
a quasar with similar redshift to image A.  The relative 
probabilities that the double is the smallest separation 
($4.8h^{-1}$ kpc for $\Omega_0=1$) correlated quasar pair or a 
gravitational lens are $\sim 1:10^6$.  If the object is a lens, the 
mean lens redshift is $\langle z_l\rangle=0.5$ with 90\% confidence 
bounds of $0.18 < z_l < 0.83$ for $\Omega_0=1$.  If the lens is an 
elliptical galaxy, we expect it to be brighter than $I < 19.5$ mag.
The broad band flux ratio varies with wavelength, with $\Delta I=0.31$, 
$\Delta R=0.58$, and $\Delta B=1.29$ magnitudes, which 
is difficult to reconcile with the lensing
hypothesis.    
\end{abstract}

\keywords{Gravitational Lenses}

\newpage

\section{Introduction}

Gravitational lenses offer the best current constraint on the cosmological
constant, with a current upper limit of $\lambda_0 < 0.65$ at 2-$\sigma$
in flat cosmological models (Kochanek 1996).  The uncertainties in the
limit are still dominated by Poisson uncertainties due to the small
number of gravitational lenses available for the analysis.  We have
been conducting a survey for gravitationally lensed quasars (Kochanek, Falco \& Schild
1995) to reduce these
uncertainties.  Here we report on the first new lens candidate found
in our sample.

SBS 0909+532 was identified in the 
course of the Hamburg-CfA Bright Quasar Survey (Engels et al. 1996). 
The quasar was discovered by Stepanyan et al. (1991), with 
coordinates accurate to $\sim 1\arcmin$ (see also V\'eron-Cetty
\& V\'eron 1996).  Our astrometric solution based on
GSC stars in the field yields $\alpha=$9:13:00.76, 
$\delta=+$52:59:31.5 (J2000), with an uncertainty of $\sim 1\arcsec$.
The object was recognized as a blue object on an
objective prism plate taken with the Calar Alto Schmidt telescope,
and subsequent spectroscopy with the Tillinghast 1.5m telescope
at FLWO (Fred Lawrence Whipple Observatory) 
on Mt. Hopkins confirmed it to be a quasar. 
SBS 0909+532 was included in the FKS 
survey for gravitationally lensed quasars (Kochanek, Falco \& Schild 1995) 
because of its redshift ($z=1.377$) and bright
optical flux ($m_B\sim17$ mag).  The quasar 
was first imaged on 19 November 1995 in 
$2\parcs9$ FWHM seeing on the 1.2m FLWO telescope on Mt. Hopkins.  The 
results were consistent with a point source,
but, because of the poor seeing conditions of the observation,
the quasar was returned to the survey observation list.  The next 
observation was on 13 January 1996 in $1\parcs2$ FWHM seeing on the 
MDM (Michigan-Dartmouth-MIT) 
2.4m telescope. The object was visibly extended, and an analysis using 
Daophot (Stetson 1992) split it into a double with a
separation of $\Delta\theta = 1\parcs1$ and flux ratio of $0.58$ $R$ mag. 

Subsequent efforts focused on determining whether both components were quasars. 
In \S2 we detail our photometric observations of the system, including 
narrow-band filters bracketing the MgII 2798 $\AA$ line of the quasar.  In 
\S3, we detail our spectroscopic observations, and in \S4 we discuss whether 
SBS 0909+532 A, B is a 
gravitational lens or the smallest known binary quasar.

\section{Photometry}

Following the identification of SBS 0909+532 as a double, on 25 January 1996 P. Schechter 
obtained multiple 60-second exposures of the quasar with the MDM 2.4m telescope
in the $I$, $R$ filters and 120-second exposures in the $B$ filter. 
The seeing conditions ranged from $0\parcs95$ to $1\parcs60$ FWHM.  
During that night, we also obtained exposures in $I$, $R$ and
$B$ of the Landolt 
photometric standard stars PG1323-086, PG1323-086A, PG1323-086C and
PG1323-086B. We used these stars to determine that conditions were 
photometric, and to calibrate the magnitudes of the quasars. 
The quasar exposures were kept short so 
that the brighter stars 
near the quasar would be unsaturated and could be used as psf 
templates. The detector
was a thinned CCD (``Charlotte'') with pixel scale $0\parcs275$, 
gain $3.45$ e$^-$/ADU, and read noise of 4e$^{-}$.  Figure 1 shows a 
$\sim 3'$ field containing SBS 0909+532, 
including the nearby stars labeled a, b, and c that we used as 
psf templates and in spectroscopic tests.  Figure 2 shows our highest
resolution $B$, $R$ and  $I$ images.  

\begin{figure}
\vspace*{120mm}
\begin{minipage}{120mm}
\includegraphics{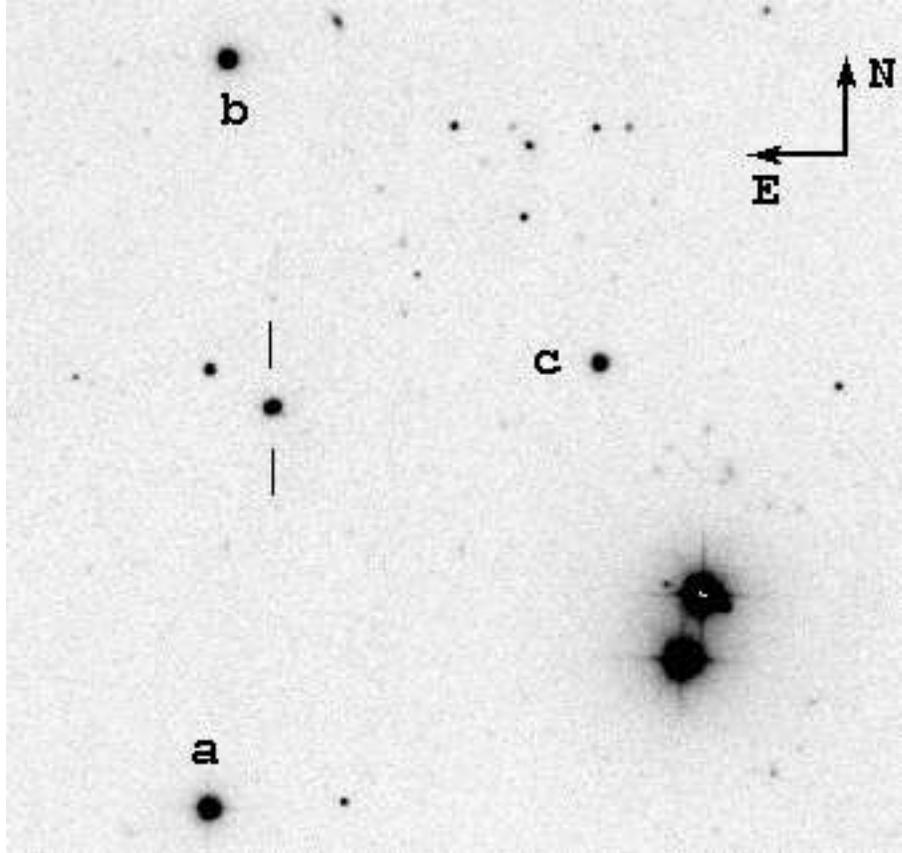}
\end{minipage}
\vskip 2.cm
\caption{Stacked I images of SBS 0909+532 A, B,
showing a $\sim 3'$ field containing the quasar,
indicated by the half-cross-hair. The
length of the arrows is 20''.}
\vskip 4.cm
\end{figure}

\begin{figure}
\vspace*{60mm}
\begin{minipage}{60mm}
\includegraphics{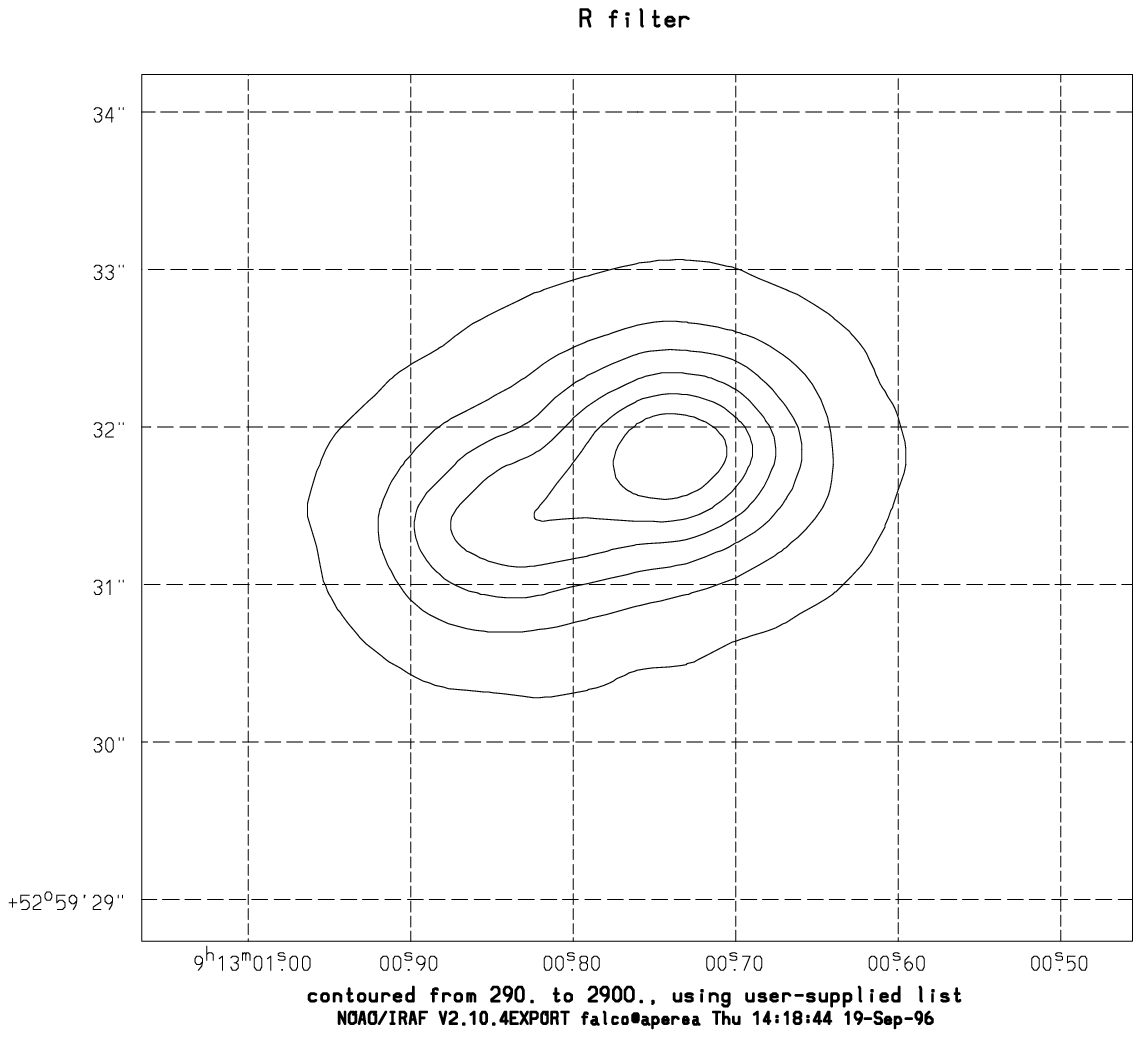}
\end{minipage}

\vspace*{60mm}
\begin{minipage}{60mm}
\includegraphics{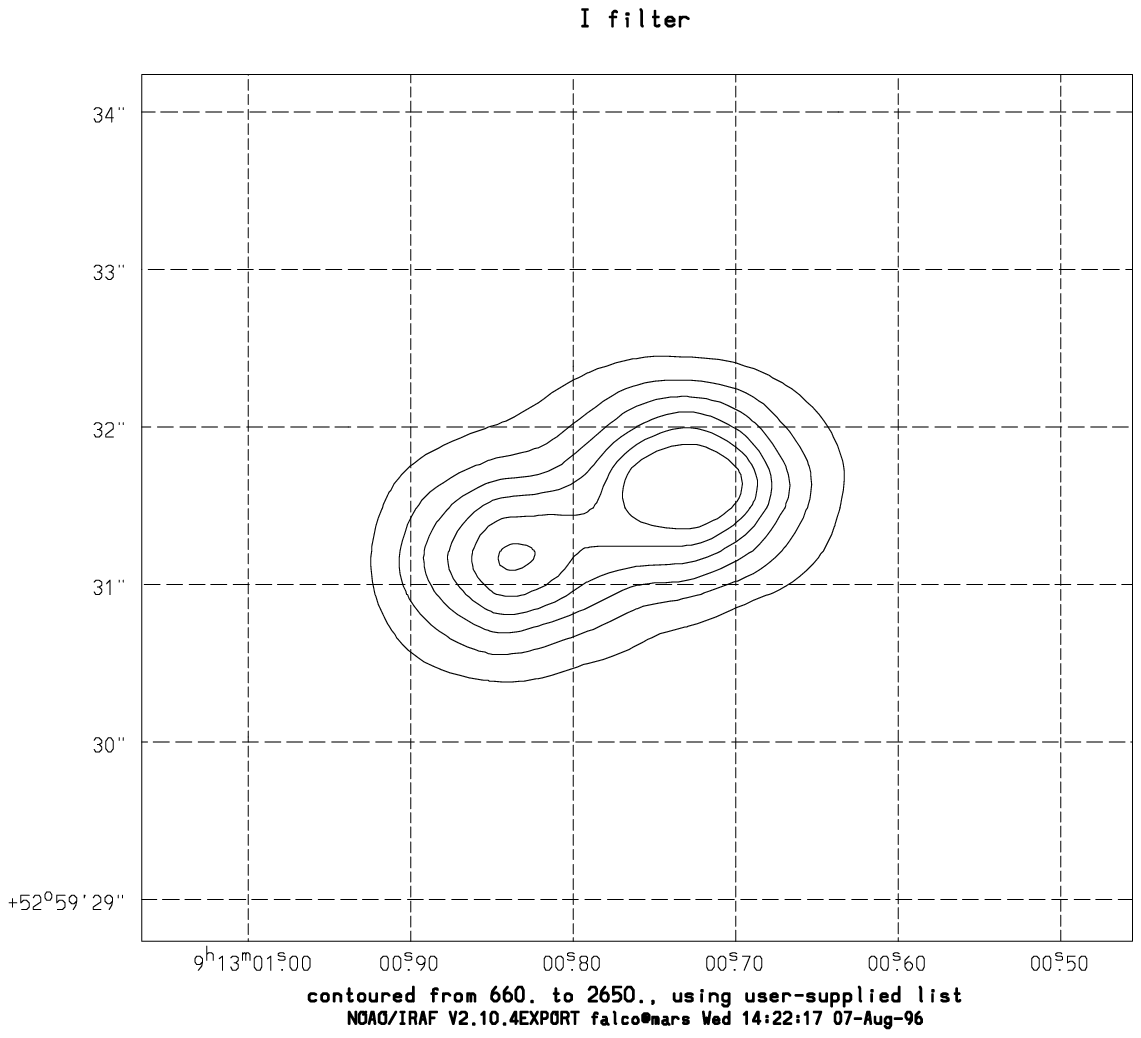}
\end{minipage}
\begin{minipage}{60mm}
\includegraphics{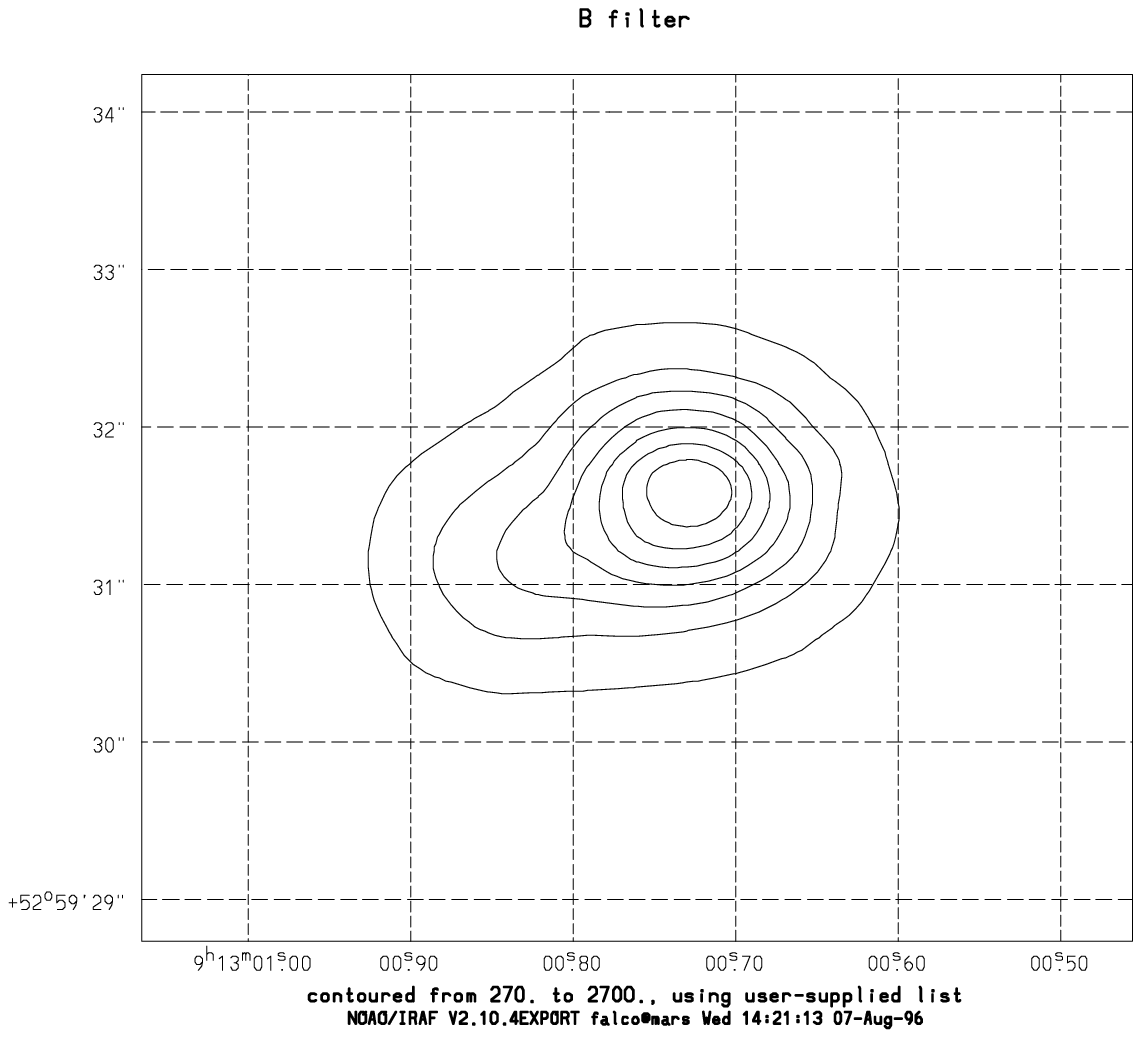}

\end{minipage}
\begin{minipage}{60mm}
\includegraphics{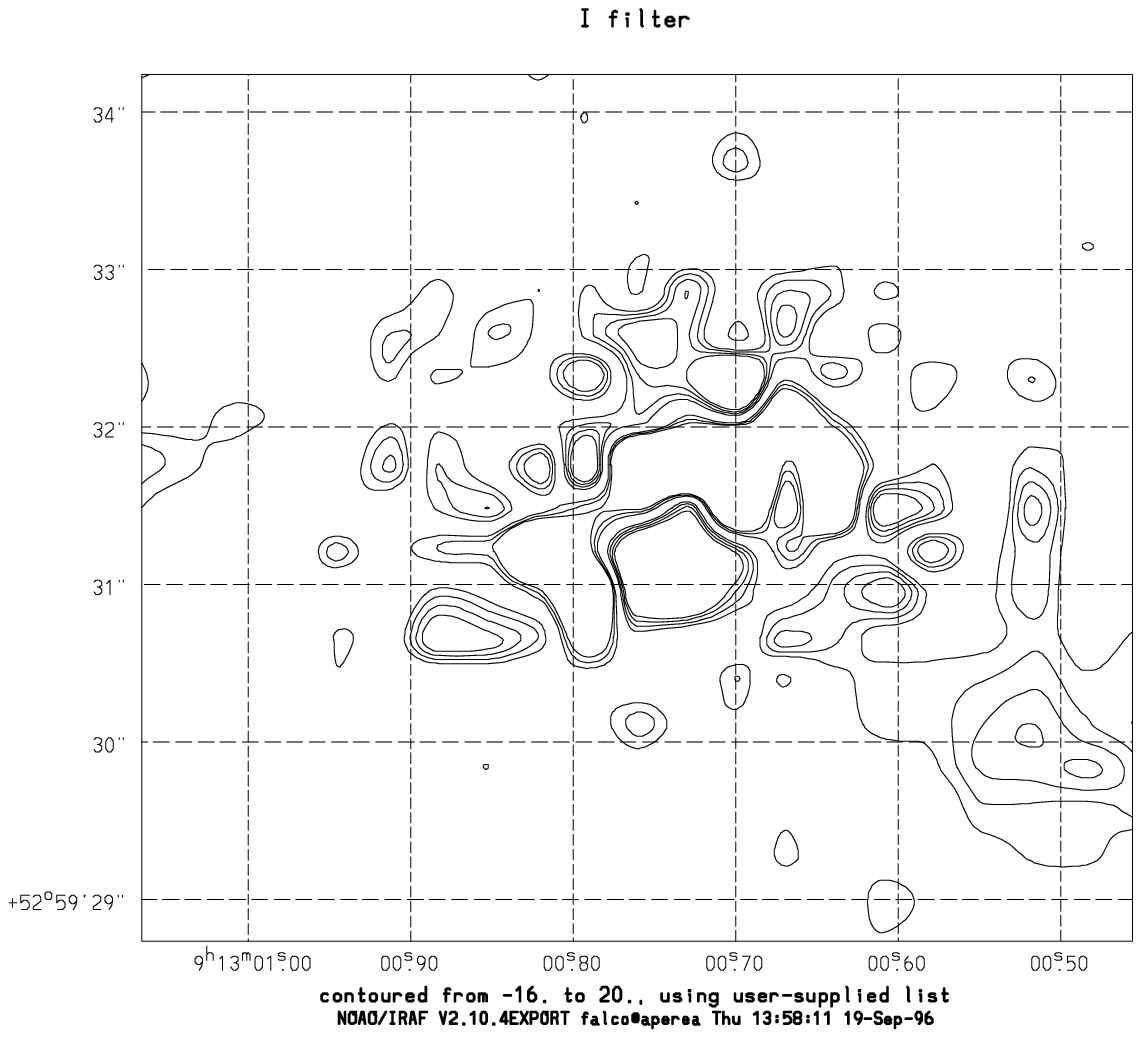}

\end{minipage}
\vskip 2.cm
\caption{Close-up views of our best-seeing $B$, $R$ and $I$
images of SBS 0909+532 A,B and of the residuals after subtraction of A and B
from the $I$ stack of Fig. 1 (clockwise from left, bottom).
The coordinates are J2000.
The A (B) component is on the right (left).
The image contours are spaced by $\sim 15$\% from the A maximum
values.
The residual contours are spaced by 2$\sigma$ between
$-4$ and $+8$ $\sigma$, where the
peak in the stacked $I$ image is $\sim 650\times\sigma$ and $\sigma$ is
the sky noise level.
The dominant features near the
locations of A and B are due to imperfect subtractions. The faint object
$\sim 2\parcs2$ SW of A and B is a possible galaxy that may be
the absorber at $z \sim 0.83$ (see text).
This object is present, but barely discernible in Fig. 1.  }
\end{figure}

Each image was processed using Daophot (Stetson 1992) 
to determine the positions and
fluxes of the two components.  
The calibrated magnitudes of the A (bright) and B (faint) components 
are listed in Table 1; their inferred separations and flux ratios
are summarized in Table 2.  We neglect galactic extinction,
which is less than $E(B-V) < 0.03$ (Burstein \& Heiles 1982).  
In all cases,  
the quasar is separable into two components, although in the worst-seeing 
$I$ image ($1\parcs6$ FWHM) the separation is not well determined.
We discard this image from the subsequent analysis.

After several attempts to obtain independent spectra of the A and B components 
failed due to poor seeing or ambiguities produced by differential refraction
(see \S3), we decided to use narrow-band filters to search for 
the MgII 2798 \AA\  emission line redshifted to 6651 \AA.  
On 11-13 April 1996 we obtained narrow-band 
images with the MDM 2.4m 
telescope in $0\parcs95$ to $1\parcs20$ FWHM seeing, in 
3 redshifted H$\alpha$ filters (see Table 3 for their properties).  
We acquired 6-minute exposures with the $KP6520$ and $KP6737$ filters,
and 10-minute exposures with the $KP6649$ filter.
The detector was a thick CCD (``Nellie'') with pixel scale $0\parcs241$, gain 
$2.95$ e$^{-1}$/ADU, and read noise of 4e$^{-}$.  Unfortunately, we had 
no comparable filter sequence for the stronger CIV 1549 \AA\  or 
CIII$\left.\right]$ 1909 \AA\ lines at 3682 \AA\  and 4538 \AA. 

We processed the narrow-band images with Daophot, in the same manner 
as the broad-band images.
The component separations (see Table 2) found for all six filters and the two 
different cameras are mutually consistent; we find a best estimate for the 
separation of $\Delta \theta = 1\parcs109 \pm 0\parcs006$ for the 
Charlotte chip and $1\parcs102\pm0\parcs007$
for the Nellie chip, where the uncertainties are the observed scatter in the
separations found for the individual images and the difference
between the two chips is probably due to 1\% errors in the chip scales.
We tested the plausibility of the uncertainties using Monte Carlo simulations.
For the best I images, the expected statistical uncertainties in the position
and flux were $0\parcs002$ and $0.001$ mag respectively, so the observed
uncertainties are dominated by systematic errors.  
The position angle of B relative to A is $115\pdeg2\pm0\pdeg3$ (E of N).

In Figure 3 we combine the calibrated total spectrum (see \S3)
with the relative
photometry to illustrate the broad features of the spectra of 
components A and B and for star c.  The MgII 2798 \AA\ line is 
clearly apparent above the $R$ continuum in both components.  The contrast 
between the on-line (KP6649) and off-line (KP6520 and KP6737) filters is 
$-0.32\pm0.05$ mag in the bright quasar component, and $-0.21\pm0.04$ mag in 
the fainter quasar component.  The continuum on either side of the line is 
flat, with a slope of $0.002\pm0.055$ mag between the KP6520 and KP6737 
filters (see Table 2), and slightly bluer than the average $R$ flux.
The equivalent width of the MgII 2798 \AA\ line 
is smaller in component B than in component A, 
because the flux ratio in the KP6649 line filter is 
$0.13\pm0.01$ mag greater than in the two off-line filters.  The broad band 
colors of component B are consistent with a K4 star, but K stars lack strong 
emission or absorption features in the wavelength range of the KP 6649 
filter.  For example, star b has the colors of a K3 star, but shows no 
line feature in the narrow band filters.  A cooler M star could 
produce features in the narrow band filters.  For example, star c is an 
M3 star, and we see the edge of one of the molecular absorption bands in 
the narrow band filters (see Figure 3).  Note, however, that we
see a gradient in the narrow band filters with the KP 6520 filter 
above the $R$ continuum, the KP 6649 filter on it, and the KP 6737 
filter below it, rather than the line feature seen in components A and B. 
We conclude that both A and B are quasars at comparable redshifts,
but that the equivalent width of the MgII 2798 \AA\ line is smaller in 
B than in A.

\section{Spectroscopy}

We made three attempts to spectroscopically resolve the two objects,
with mixed success due to poor seeing during each of the observations.
We report only on the best of these observations.
On 22 April 1996, we obtained long slit spectra of the quasar with a
$1'' \times 180''$ slit parallel to the component separation. 
We used the MMT with the Blue Channel spectrograph and a Loral 3K$\times$1K
CCD.  With the same slit in the same orientation, we obtained spectra of 
star c just before and after the spectrum of the quasar.  The spatial 
scale of the image is $0\parcs30$ per pixel, the wavelength scale is 
$1.96$ \AA/pixel, the gain is $1.5$ and the read noise $7.5$ $e^-$.  
The seeing during the observations was $1\parcs9$ FWHM; the spectral 
resolution was $\sim 6.5\,$\AA\ FWHM. 
In the resulting spectrum, the two components of the SBS 0909+532 system are 
not separable by standard reduction methods. 
On 20 April 1996, we had obtained spectra of the quasar with the 
same instrumental setup, but the seeing was significantly worse. 
That night, we did
obtain spectra of the Landolt standard PG1121+145, which we used to
derive an approximate photometric calibration for our quasar spectra.
Figure 4 shows the 
calibrated spectra of the combined A/B components and of star c.

We identified a series of heavy-element absorption lines in the
quasar spectrum that correspond well with C III$\left.\right]$, Fe II
and Mg II at $z \approx 0.83$ (see Figure 4). Table 4 lists the 
identified lines with their wavelengths and estimated equivalent widths. 
The equivalent width is typical of Mg II absorption systems (Steidel
\& Sargent 1992) over a wide range of redshifts and impact 
parameters (e.g. Steidel, Dickinson \& Persson 1994, 
Churchill, Steidel \& Vogt 1996). While the absorption redshift is
a plausible lens redshift (see \S4), such absorption systems
are common in unlensed quasars and we cannot regard it as 
evidence in support of the lensing hypothesis.  Steidel et al. (1994)
and Churchill et al.  (1996) 
have found, however, that the galaxies responsible for
Mg II absorption systems are usually detectable, and in our
images there is a plausible, faint candidate galaxy $2\parcs2$
SW of component A with $m_I \sim 21$ (see Figure 2).  
The candidate galaxy is not
clearly observed in the $B$ or $R$ images.
The implied impact parameter of $9h^{-1}$ kpc is typical of
Mg II absorption system galaxies with the observed equivalent 
width.

We also combined the spectra of the similar redshift quasars 1257+3439,
1356+5806, and 2340+0019 (Steidel \& Sargent 1992) with spectra
of a K4 dwarf and a K4 giant star scaled to the observed continuum ratios
(the results are insensitive to variations of spectral type between
K3 and K5).
The stars have a broad MgIb absorption feature near
5000-5300 \AA\  which is prominent even when added to the 
spectrum of the quasar.  The absence of the feature in the
observed spectrum (Figure 4) is further evidence that component B
cannot be a star.

\begin{figure}
\centerline{\psfig{figure=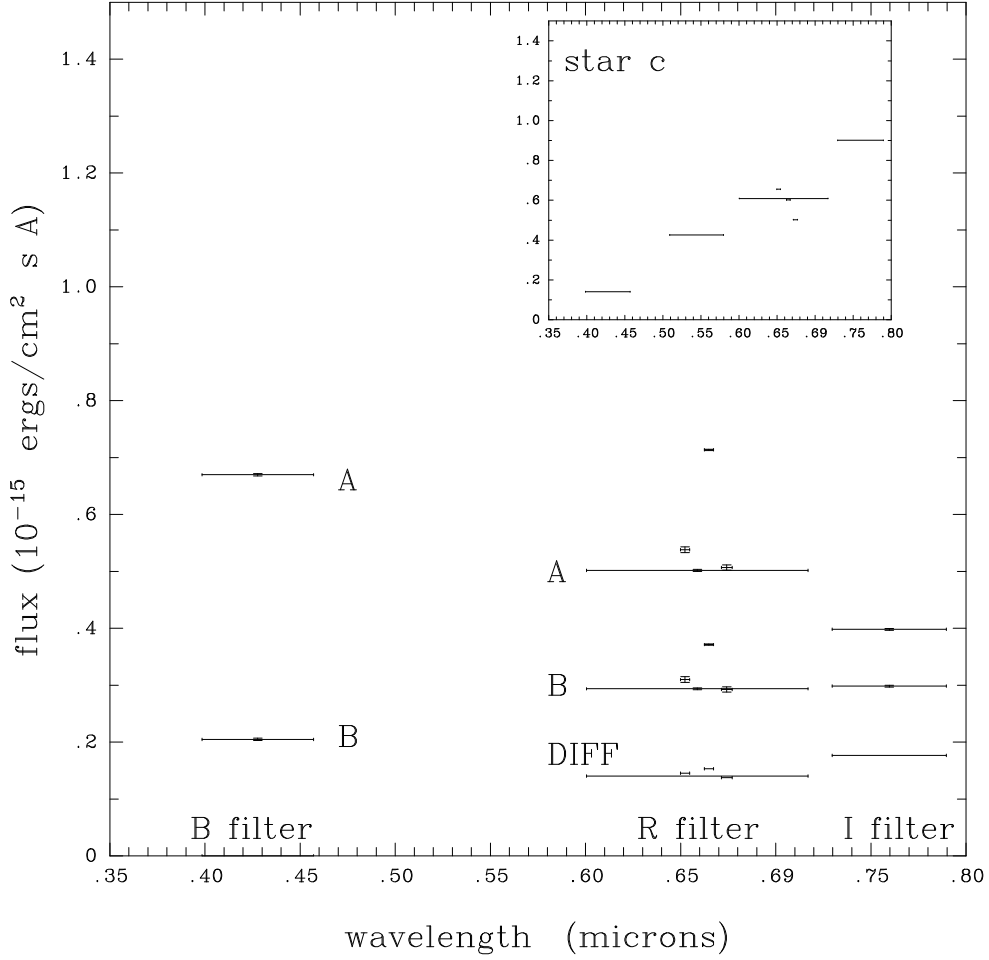,width=5.0in}}
\caption{Narrow-band photometry of the separate A and B components
  estimated from the calibrated spectrum and the photometry.  The
  inset shows the analogous measurements for star c.  The error
  bars in wavelength show the rms width of the filter, and the
  error bars in flux represent the uncertainty in the relative
  fluxes from the photometry.  The DIFF values show the residuals
  after scaling the A component photometry
   to the $B$-filter flux of component B
  and subtracting. The $I$ filter averages are truncated before the red
  edge of the passband because the calibrated spectrum ends at 8000 \AA. }
\end{figure}

\begin{figure}
\centerline{\psfig{figure=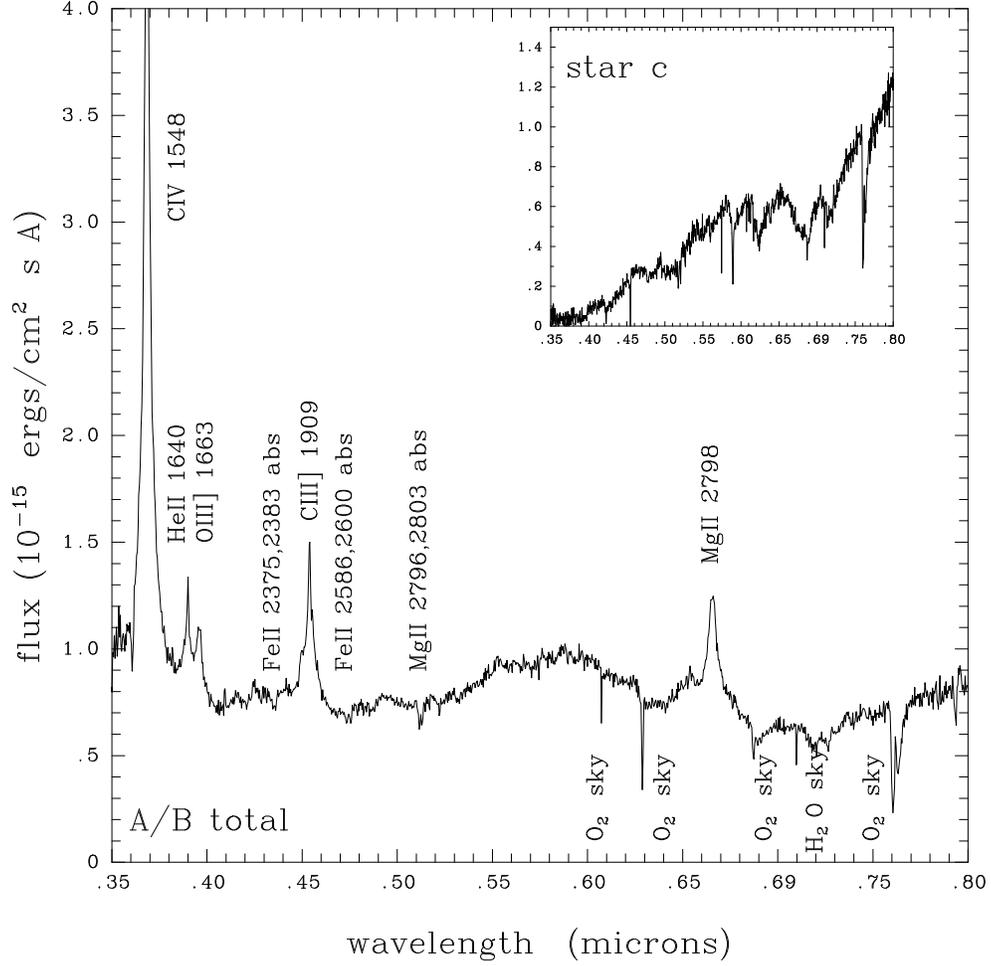,width=5.0in}}
\caption{The calibrated spectrum of the combined A/B components.  The
  inset shows the calibrated spectrum of star c. The $z_{abs}=1.33$
  C IV doublet lies just inside the left edge.  }
\end{figure}

\begin{figure}
\centerline{\psfig{figure=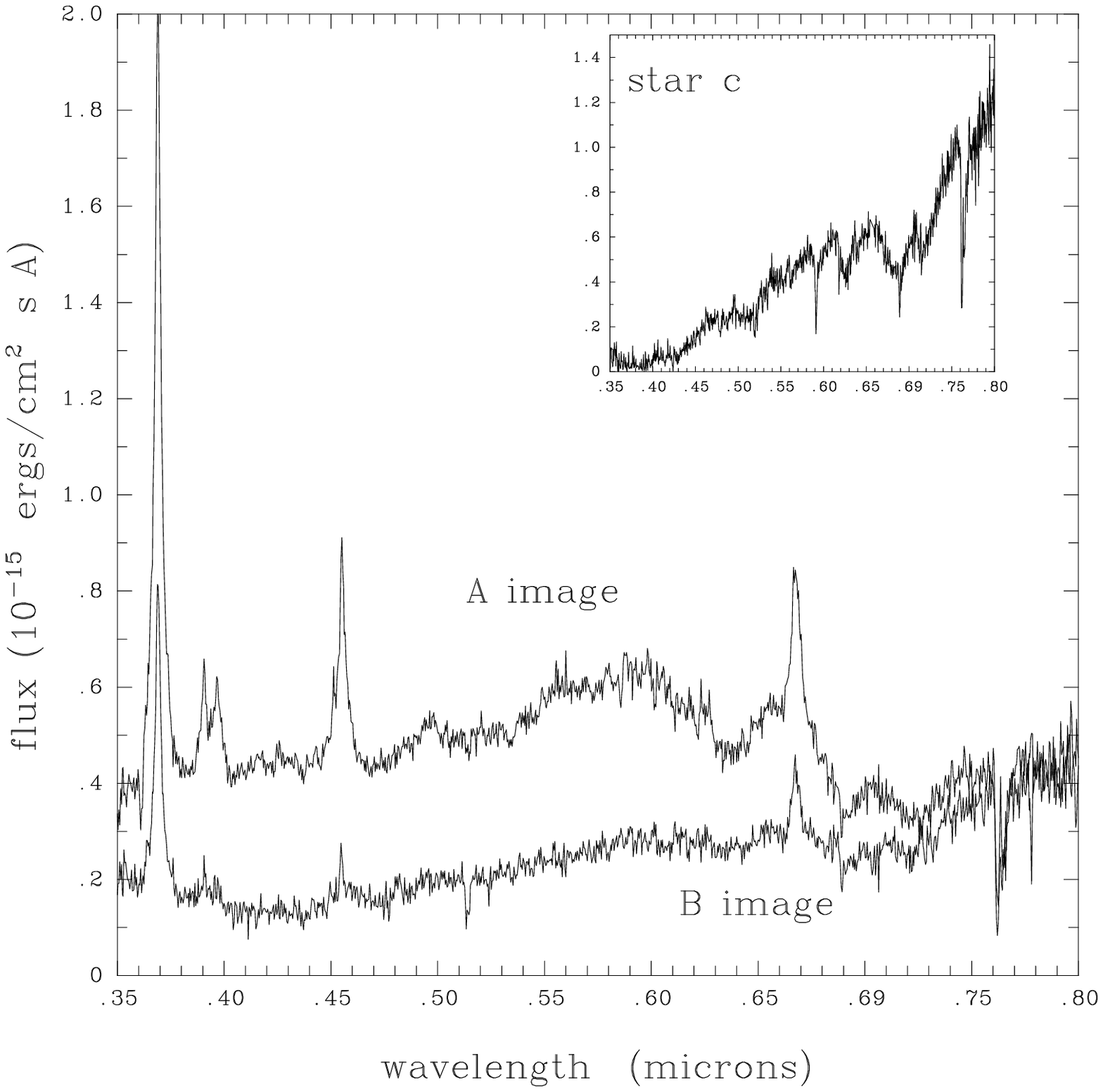,width=5.0in}}
\caption{Estimated spectra of the A and B components.  The spectra are
  most reliable near the Mg II line because of the additional
  constraints from the narrow band photometry.  There are
  significant unmodeled residuals in the region with the poorly 
  split OIII$\left.\right]$ and CIII$\left.\right]$ emission lines and MgII
  absorption line.
   }
\end{figure}

The poor seeing at the time we obtained the spectrum makes it
very difficult to separate the two objects.  We attempted to 
do so by modeling the spatially resolved data with a flat 
background and two positive definite spectra with the 
photometrically-determined separation, convolved with a Gaussian
seeing profile.  Experiments with the spectrum of star c 
demonstrated that our model would extract a spectrum matching
the one obtained by standard methods shown in Figure 3, and
experiments with Monte Carlo simulations of the quasar 
components demonstrated that the extraction procedure would not
seriously contaminate the emission line properties of the
two components given the separation, seeing, and noise properties
of the data.  We forced the extracted spectra to match the
observed $B$, $R$, and narrow band flux ratios to reduce the
ambiguities in the solution.  The residuals for fitting the
A/B components are not as good as the residuals when we modeled
the spectrum of star c, particularly in the $V$ band region where
we lacked photometric constraints.  
Figure 5 shows the resulting spectra for A and B.  The results
are most reliable in the region near the narrow band filters,
moderately reliable in the region near the broad band $B$ and $R$ 
filters, and of dubious reliability elsewhere.  The 
region from the red edge of the B band through the V band
is particularly problematic, since the HeII, OIII$\left.\right]$,
and CIII$\left.\right]$ lines appear mainly in the estimated spectrum
of A, the MgII absorption feature appears mainly in
the spectrum of B, and there are significant unmodeled
residuals.

\section{Interpretation}

We believe that component B is not a star for three reasons.
First, the MgII 2798 \AA\ line is present in both components based
either on the narrow band photometry or the spectral 
reconstruction.  Second, the spectral reconstruction, 
while not ideal, generally required emission lines in both
objects, particularly in the regions where the model was
stabilized by the broad band photometry.  Third, K stars 
should produce a broad absorption feature in the combined
(A $+$ B) spectrum that we do not observe.
If both objects possess broad emission lines with comparable 
redshifts, the two plausible explanations are that the object is a 
gravitational lens or that it is the smallest-separation binary  
quasar yet found. 

Using the procedures of Kochanek (1992) we estimate that the 
relative probabilities of finding two lensed or correlated quasars with 
the observed separation and magnitude difference are approximately 
$p_{lens} \sim 0.02$ and $p_{corr} \sim 10^{-8}$.  The
probability that the system is a lens is overwhelmingly favored because 
quasars as bright as SBS 0909+532 A, B have very low space densities.  
Although the velocity difference is consistent with zero
within our estimates for random errors, we cannot
sensibly derive any limits due to the 
systematic effects on our estimates of 
the separate spectra in \S3.  The projected separation of the
two quasars, if they are distinct objects, is only $4.7 h^{-1}$ kpc 
(for $\Omega_0=1$).  All other multiple quasar
systems with such small separations are lenses, while the smallest 
binary quasar known
not to be a lens is PKS 1145--071 ($\Delta \theta =4\parcs2$, 
$\Delta m=0.8$, $z=1.35$,
Djorgovski et al. 1987) with a projected separation of $18h^{-1}$ kpc.  
The broad band 
color differences are, however, only easily explained for a 
binary quasar.

\begin{figure}
\centerline{\psfig{figure=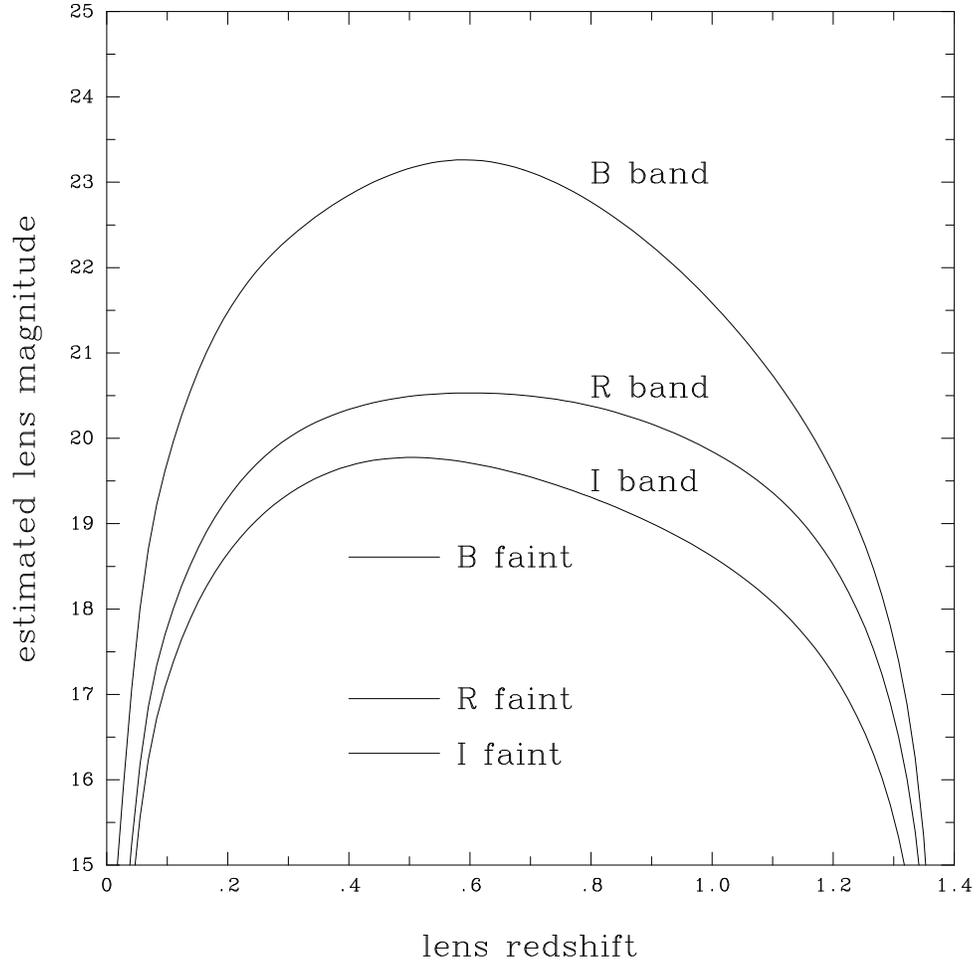,height=5.0in}}
\caption{Estimated magnitude of a lens galaxy as a function of
the lens redshift, assuming the lens is an elliptical galaxy.
Horizontal bars indicate the magnitude of the faint component in
each broad band filter.}
\end{figure}

The only obstacles to explaining the SBS 0909+532 A/B system as a 
lensed pair are differences in the spectra of the two objects.  In 
a gravitational lens, spectral differences can be
produced by time delays, microlensing, 
emission by the lens galaxy, and 
extinction by the lens galaxy.  Time delays and microlensing are 
unlikely to produce a difference in the spectra of the 
breadth in wavelength and amplitude of flux that is observed as an apparent
reddening of B. For a standard extinction curve to produce the 
measured spectral differences, component B must have $E(B-V) \simeq 0.43$ more
extinction than component A, and the intrinsic flux ratio must be 
$-0.58$ mags (B is intrinsically brighter than A).  If the spectral
differences are due to the presence of a lens galaxy, then we expect the blue
flux ratio to be the best estimate of the true flux ratio because
it is dominated by the broad CIV (1549 \AA) emission line.  If we scale the
photometry of A to match the $B$-filter flux of component B, and subtract,
the residuals no longer have a CIV line feature in the narrow
band photometry (see Figure 4). In any other scenario (B not a quasar, flux
differences due to reddening, two different quasars) such a 
cancellation would be a remarkable chance coincidence.  The 
estimated magnitude of the galaxy is then $I \sim 17$ mag, and the 
flux of the galaxy is a substantial fraction of the total
flux attributed to component B.  However, Monte Carlo simulations show
that such a bright galaxy would produce detectable residuals in 
our data.

If we assume an SIS lens model, we can estimate the lens redshift 
and magnitude using the procedures 
of Kochanek (1992).  For $\Omega=1$ the mean redshift 
is $\langle z_l \rangle = 0.50 \pm 0.20$, and 
90\% of the probability lies between $ 0.18 < z_l < 0.83$. 
 The luminosity of the lens galaxy is 
$L/L_* = (\Delta\theta/\Delta\theta_* (1-x))^{2/\gamma}$ where 
$x=D_{LS}/D_{OS}$, $D_{LS}$ and $D_{OS}$ are angular-size 
distances from the lens and the observer to the source, 
respectively, 
$\Delta\theta_* = 2\parcs92(\sigma_*/225\kms)^2$ is the splitting 
scale for an $L_*$
galaxy, $\sigma_*$ is
the corresponding velocity dispersion, 
and $\gamma\simeq 4$ is a ``Tully-Fisher'' exponent.  
Figure 6 shows the
expected magnitude of the lens galaxy assuming an $L_*$ elliptical 
galaxy has absolute
luminosity $B_* = -19.9+5\log h$ mag and the K-corrections and 
evolution models of Guiderdoni \& Rocca-Volmerange (1988).  
A galaxy with  $I \simeq 17$ is also predicted to have a lens redshift
of $z_l \simeq 0.1$, significantly below the 90\% confidence range
for the lens redshift.

Our case for SBS 0909+532 remains somewhat circumstantial due
to the limitations of the data.  Refining our results requires
spectra in the good seeing conditions that consistently 
eluded us, and higher-resolution images.  The identification of a 
lens may be the simplest means of resolving the question, because
the lens interpretation appears to require a bright lens galaxy.
We intend to carry out infrared observations to search for such 
a galaxy to take advantage of both better imaging conditions, 
and reduced quasar-galaxy contrast.

\acknowledgments 
Acknowledgements: We thank Paul Schechter and Craig Foltz for 
helping us obtain data on the SBS 0909+532 system.  We also thank the KPNO 
staff, particularly Ed Carder, for lending us the narrow band filters used 
in \S2.  We thank B. McLeod for checking several photometric results using
methods independent of Daophot, and C. Keeton for checking the evolution
model.  A. Dobrzycki is supported by NASA contract NAS8-39073.
The Hamburg Bright Quasar Survey is 
supported by the DFG through grants Re 353/11 and Re 353/22.

\clearpage

\begin{table}[t]
\begin{center}
\begin{tabular}{rccccc}
\multicolumn{6}{c}{Table 1: Absolute Photometry} \\
\hline
Filter &$m_A$ &$m_B$ &$m_a$ &$m_b$ &$m_c$ \\ 
\hline
$I$       &$16.00\pm0.03$  &$16.31\pm0.05$ &$13.54\pm0.11$ &$14.22\pm0.04$ &$15.05\pm0.09$ \\
$R$       &$16.36\pm0.04$  &$16.95\pm0.04$ &$13.94\pm0.12$ &$14.78\pm0.05$ &$16.19\pm0.05$ \\
$B$       &$17.24\pm0.03$  &$18.61\pm0.03$ &$14.79\pm0.08$ &$16.02\pm0.02$ &$18.68\pm0.03$ \\
\hline
\end{tabular}
\end{center}
Notes -- A and B are the quasar images, and a, b, and c are the reference
stars (see Figure 1). The absolute photometry is significantly less 
accurate than the relative photometry presented in Table 2.  The broad band 
filters are the Kitt Peak $I$ and $B$ filters, and the Schombert $R$ filter.
\end{table}

\clearpage

\begin{table}[t]
\begin{center}
\begin{tabular}{ccccc}
\multicolumn{5}{c}{Table 2: Relative Photometry} \\
\hline
Filter &$N_{im}$ &$\Delta\theta$ &$\Delta m$ &$\Delta m_{AB}$ \\
\hline
$I$      &9 &$1\parcs107\pm0\parcs006$ &$0.313\pm0.010$   &--  \\
$R$      &8 &$1\parcs109\pm0\parcs005$ &$0.581\pm0.010$   &$\equiv0$ \\
$B$      &4 &$1\parcs111\pm0\parcs006$ &$1.286\pm0.015$   &--        \\
$KP6520$ &6 &$1\parcs104\pm0\parcs006$ &$0.599\pm0.028$   &$0.018\pm0.030$ \\
$KP6649$ &3 &$1\parcs102\pm0\parcs001$ &$0.709\pm0.005$   &$0.128\pm0.011$ \\
$KP6737$ &3 &$1\parcs101\pm0\parcs007$ &$0.596\pm0.027$   &$0.015\pm0.028$ \\
\hline
\end{tabular}
\end{center}
Notes -- $N_{im}$ is the number of images used in the analysis; it
includes only images with FWHM better than $1\parcs2$.  The errors in the
separation $\Delta\theta$ and the flux ratios $\Delta m$ are the observed
dispersion between the images.  
$\Delta m_{AB}=(m_A-m_B)-(R_A-R_B)$ is the color difference between 
the quasar images in each band compared to the quasar images in $R$.  
The position angle of B relative to A is $115\pdeg2\pm0\pdeg3$ (E of N).
\end{table}

\clearpage

\begin{table}[t]
\begin{center}
\begin{tabular}{lccc}
\multicolumn{4}{c}{Table 3: Properties of Narrow-band Filters} \\
\hline
Filter &$\lambda_c$ & FWHM $\Delta\lambda$ & Peak transmission\\
 &         (\AA) & (\AA) & (\%)\\
\hline
$KP6520$       & 6524 & 70 & 84\\
$KP6649$       & 6649& 71 & 87\\
$KP6737$       & 6743& 84 & 89\\
\hline
\end{tabular}
\end{center}
\end{table}

\clearpage

\begin{table}[t]
\begin{center}
\begin{tabular}{lccc}
\multicolumn{4}{c}{Table 4: Absorption Lines} \\
\hline
Element &rest $\lambda$ & rest $W_\circ$ &$z$ \\
 &         (\AA) & (\AA) & \\
\hline
C IV       & 1549 & 2.4   & 1.33\phantom{0}\\
C III $\left.\right]$  &1909 & 1.4 & 0.833 \\
Fe II & 2375 & 1.0 & 0.831 \\
Fe II & 2383 & 0.3 & 0.830 \\
Fe II & 2587 & 0.9 & 0.831 \\
Fe II & 2600 & 0.5 & 0.829 \\
Mg II & 2796 & 1.0 & 0.830 \\
Mg II & 2803 & 0.8 & 0.830 \\
\hline
\end{tabular}
\end{center}
\end{table}

\end{document}

%% file: psfig2.tex
\def\PsfigVersion{1.10}
\def\setDriver{\DvipsDriver} 
\ifx\undefined\psfig\else \fi
\let\LaTeXAtSign=\@
\let\@=\relax
\edef\psfigRestoreAt{\catcode`\@=\number\catcode`@\relax}
\catcode`\@=11\relax
\newwrite\@unused
\def\ps@typeout#1{{\let\protect\string\immediate\write\@unused{#1}}}

\def\DvipsDriver{
	\ps@typeout{psfig/tex \PsfigVersion -dvips}
\def\PsfigSpecials{\DvipsSpecials} 	\def\ps@dir{/}
\def\ps@predir{} }
\def\OzTeXDriver{
	\ps@typeout{psfig/tex \PsfigVersion -oztex}
	\def\PsfigSpecials{\OzTeXSpecials}
	\def\ps@dir{:}
	\def\ps@predir{:}
	\catcode`\^^J=5
}

\def\figurepath{./:}

\def\DoPaths#1{\expandafter\EachPath#1\stoplist}
\def\leer{}
\def\EachPath#1:#2\stoplist{
  \ExistsFile{#1}{\SearchedFile}
  \ifx#2\leer
  \else
    \expandafter\EachPath#2\stoplist
  \fi}
\def\ps@dir{/}
\def\ExistsFile#1#2{%
   \openin1=\ps@predir#1\ps@dir#2
   \ifeof1
       \closein1
   \else
       \closein1
        \ifx\ps@founddir\leer
           \edef\ps@founddir{#1}
        \fi
   \fi}
\def\get@dir#1{%
  \def\ps@founddir{}
  \def\SearchedFile{#1}
  \DoPaths\figurepath
}

\def\@nnil{\@nil}
\def\@empty{}
\def\@psdonoop#1\@@#2#3{}
\def\@psdo#1:=#2\do#3{\edef\@psdotmp{#2}\ifx\@psdotmp\@empty \else
    \expandafter\@psdoloop#2,\@nil,\@nil\@@#1{#3}\fi}
\def\@psdoloop#1,#2,#3\@@#4#5{\def#4{#1}\ifx #4\@nnil \else
       #5\def#4{#2}\ifx #4\@nnil \else#5\@ipsdoloop #3\@@#4{#5}\fi\fi}
\def\@ipsdoloop#1,#2\@@#3#4{\def#3{#1}\ifx #3\@nnil 
       \let\@nextwhile=\@psdonoop \else
      #4\relax\let\@nextwhile=\@ipsdoloop\fi\@nextwhile#2\@@#3{#4}}
\def\@tpsdo#1:=#2\do#3{\xdef\@psdotmp{#2}\ifx\@psdotmp\@empty \else
    \@tpsdoloop#2\@nil\@nil\@@#1{#3}\fi}
\def\@tpsdoloop#1#2\@@#3#4{\def#3{#1}\ifx #3\@nnil 
       \let\@nextwhile=\@psdonoop \else
      #4\relax\let\@nextwhile=\@tpsdoloop\fi\@nextwhile#2\@@#3{#4}}
\ifx\undefined\fbox
\newdimen\fboxrule
\newdimen\fboxsep
\newdimen\ps@tempdima
\newbox\ps@tempboxa
\long\def\fbox#1{\leavevmode\setbox\ps@tempboxa\hbox{#1}\ps@tempdima\fboxrule
    \advance\ps@tempdima \fboxsep \advance\ps@tempdima \dp\ps@tempboxa
   \hbox{\lower \ps@tempdima\hbox
  {\vbox{\hrule height \fboxrule
          \hbox{\vrule width \fboxrule \hskip\fboxsep
          \vbox{\vskip\fboxsep \box\ps@tempboxa\vskip\fboxsep}\hskip 
                 \fboxsep\vrule width \fboxrule}
                 \hrule height \fboxrule}}}}
\fi
\newread\ps@stream
\newif\ifnot@eof       
\newif\if@noisy        
\newif\if@atend        
\newif\if@psfile       
{\catcode`\%=12\global\gdef\epsf@start{
\def\epsf@PS{PS}
\def\epsf@getbb#1{%
\openin\ps@stream=\ps@predir#1
\ifeof\ps@stream\ps@typeout{Error, File #1 not found}\else
   {\not@eoftrue \chardef\other=12
    \def\do##1{\catcode`##1=\other}\dospecials \catcode`\ =10
    \loop
       \if@psfile
	  \read\ps@stream to \epsf@fileline
       \else{
	  \obeyspaces
          \read\ps@stream to \epsf@tmp\global\let\epsf@fileline\epsf@tmp}
       \fi
       \ifeof\ps@stream\not@eoffalse\else
       \if@psfile\else
       \expandafter\epsf@test\epsf@fileline:. \\%
       \fi
          \expandafter\epsf@aux\epsf@fileline:. \\%
       \fi
   \ifnot@eof\repeat
   }\closein\ps@stream\fi}%
\long\def\epsf@test#1#2#3:#4\\{\def\epsf@testit{#1#2}
			\ifx\epsf@testit\epsf@start\else
\ps@typeout{Warning! File does not start with `\epsf@start'.  It may not be a PostScript file.}
			\fi
			\@psfiletrue} 
{\catcode`\%=12\global\let\epsf@percent=
\long\def\epsf@aux#1#2:#3\\{\ifx#1\epsf@percent
   \def\epsf@testit{#2}\ifx\epsf@testit\epsf@bblit
	\@atendfalse
        \epsf@atend #3 . \\%
	\if@atend	
	   \if@verbose{
		\ps@typeout{psfig: found `(atend)'; continuing search}
	   }\fi
        \else
        \epsf@grab #3 . . . \\%
        \not@eoffalse
        \global\no@bbfalse
        \fi
   \fi\fi}%
\def\epsf@grab #1 #2 #3 #4 #5\\{%
   \global\def\epsf@llx{#1}\ifx\epsf@llx\empty
      \epsf@grab #2 #3 #4 #5 .\\\else
   \global\def\epsf@lly{#2}%
   \global\def\epsf@urx{#3}\global\def\epsf@ury{#4}\fi}%
\def\epsf@atendlit{(atend)} 
\def\epsf@atend #1 #2 #3\\{%
   \def\epsf@tmp{#1}\ifx\epsf@tmp\empty
      \epsf@atend #2 #3 .\\\else
   \ifx\epsf@tmp\epsf@atendlit\@atendtrue\fi\fi}

\chardef\psletter = 11 
\chardef\other = 12

\newif \ifdebug 
\newif\ifc@mpute 
\c@mputetrue 

\let\then = \relax
\def\r@dian{pt }
\let\r@dians = \r@dian
\let\dimensionless@nit = \r@dian
\let\dimensionless@nits = \dimensionless@nit
\def\internal@nit{sp }
\let\internal@nits = \internal@nit
\newif\ifstillc@nverging
\def \Mess@ge #1{\ifdebug \then \message {#1} \fi}

{ 
	\catcode `\@ = \psletter
	\gdef \nodimen {\expandafter \n@dimen \the \dimen}
	\gdef \term #1 #2 #3%
	       {\edef \t@ {\the #1}
		\edef \t@@ {\expandafter \n@dimen \the #2\r@dian}%
		\t@rm {\t@} {\t@@} {#3}%
	       }
	\gdef \t@rm #1 #2 #3%
	       {{%
		\count 0 = 0
		\dimen 0 = 1 \dimensionless@nit
		\dimen 2 = #2\relax
		\Mess@ge {Calculating term #1 of \nodimen 2}%
		\loop
		\ifnum	\count 0 < #1
		\then	\advance \count 0 by 1
			\Mess@ge {Iteration \the \count 0 \space}%
			\Multiply \dimen 0 by {\dimen 2}%
			\Mess@ge {After multiplication, term = \nodimen 0}%
			\Divide \dimen 0 by {\count 0}%
			\Mess@ge {After division, term = \nodimen 0}%
		\repeat
		\Mess@ge {Final value for term #1 of 
				\nodimen 2 \space is \nodimen 0}%
		\xdef \Term {#3 = \nodimen 0 \r@dians}%
		\aftergroup \Term
	       }}
	\catcode `\p = \other
	\catcode `\t = \other
	\gdef \n@dimen #1pt{#1} 
}

\def \Divide #1by #2{\divide #1 by #2} 

\def \Multiply #1by #2
       {{
	\count 0 = #1\relax
	\count 2 = #2\relax
	\count 4 = 65536
	\Mess@ge {Before scaling, count 0 = \the \count 0 \space and
			count 2 = \the \count 2}%
	\ifnum	\count 0 > 32767 
	\then	\divide \count 0 by 4
		\divide \count 4 by 4
	\else	\ifnum	\count 0 < -32767
		\then	\divide \count 0 by 4
			\divide \count 4 by 4
		\else
		\fi
	\fi
	\ifnum	\count 2 > 32767 
	\then	\divide \count 2 by 4
		\divide \count 4 by 4
	\else	\ifnum	\count 2 < -32767
		\then	\divide \count 2 by 4
			\divide \count 4 by 4
		\else
		\fi
	\fi
	\multiply \count 0 by \count 2
	\divide \count 0 by \count 4
	\xdef \product {#1 = \the \count 0 \internal@nits}%
	\aftergroup \product
       }}

\def\r@duce{\ifdim\dimen0 > 90\r@dian \then   
		\multiply\dimen0 by -1
		\advance\dimen0 by 180\r@dian
		\r@duce
	    \else \ifdim\dimen0 < -90\r@dian \then  
		\advance\dimen0 by 360\r@dian
		\r@duce
		\fi
	    \fi}

\def\Sine#1%
       {{%
	\dimen 0 = #1 \r@dian
	\r@duce
	\ifdim\dimen0 = -90\r@dian \then
	   \dimen4 = -1\r@dian
	   \c@mputefalse
	\fi
	\ifdim\dimen0 = 90\r@dian \then
	   \dimen4 = 1\r@dian
	   \c@mputefalse
	\fi
	\ifdim\dimen0 = 0\r@dian \then
	   \dimen4 = 0\r@dian
	   \c@mputefalse
	\fi
	\ifc@mpute \then
		\divide\dimen0 by 180
		\dimen0=3.141592654\dimen0
		\dimen 2 = 3.1415926535897963\r@dian 
		\divide\dimen 2 by 2 
		\Mess@ge {Sin: calculating Sin of \nodimen 0}%
		\count 0 = 1 
		\dimen 2 = 1 \r@dian 
		\dimen 4 = 0 \r@dian 
		\loop
			\ifnum	\dimen 2 = 0 
			\then	\stillc@nvergingfalse 
			\else	\stillc@nvergingtrue
			\fi
			\ifstillc@nverging 
			\then	\term {\count 0} {\dimen 0} {\dimen 2}%
				\advance \count 0 by 2
				\count 2 = \count 0
				\divide \count 2 by 2
				\ifodd	\count 2 
				\then	\advance \dimen 4 by \dimen 2
				\else	\advance \dimen 4 by -\dimen 2
				\fi
		\repeat
	\fi		
			\xdef \sine {\nodimen 4}%
       }}

\def\Cosine#1{\ifx\sine\UnDefined\edef\Savesine{\relax}\else
		             \edef\Savesine{\sine}\fi
	{\dimen0=#1\r@dian\advance\dimen0 by 90\r@dian
	 \Sine{\nodimen 0}
	 \xdef\cosine{\sine}
	 \xdef\sine{\Savesine}}}	      

\def\psdraft{
	\def\@psdraft{0}
}
\def\psfull{
	\def\@psdraft{100}
}

\psfull

\newif\if@scalefirst
\def\psscalefirst{\@scalefirsttrue}
\def\psrotatefirst{\@scalefirstfalse}
\psrotatefirst

\newif\if@draftbox
\def\psnodraftbox{
	\@draftboxfalse
}
\def\psdraftbox{
	\@draftboxtrue
}
\@draftboxtrue

\newif\if@prologfile
\newif\if@postlogfile
\def\pssilent{
	\@noisyfalse
}
\def\psnoisy{
	\@noisytrue
}
\psnoisy
\newif\if@bbllx
\newif\if@bblly
\newif\if@bburx
\newif\if@bbury
\newif\if@height
\newif\if@width
\newif\if@rheight
\newif\if@rwidth
\newif\if@angle
\newif\if@clip
\newif\if@verbose
\def\@p@@sclip#1{\@cliptrue}
\newif\if@decmpr
\def\@p@@sfigure#1{\def\@p@sfile{null}\def\@p@sbbfile{null}\@decmprfalse
   \openin1=\ps@predir#1
   \ifeof1
	\closein1
	\get@dir{#1}
	\ifx\ps@founddir\leer
		\openin1=\ps@predir#1.bb
		\ifeof1
			\closein1
			\get@dir{#1.bb}
			\ifx\ps@founddir\leer
				\ps@typeout{Can't find #1 in \figurepath}
			\else
				\@decmprtrue
				\def\@p@sfile{\ps@founddir\ps@dir#1}
				\def\@p@sbbfile{\ps@founddir\ps@dir#1.bb}
			\fi
		\else
			\closein1
			\@decmprtrue
			\def\@p@sfile{#1}
			\def\@p@sbbfile{#1.bb}
		\fi
	\else
		\def\@p@sfile{\ps@founddir\ps@dir#1}
		\def\@p@sbbfile{\ps@founddir\ps@dir#1}
	\fi
   \else
	\closein1
	\def\@p@sfile{#1}
	\def\@p@sbbfile{#1}
   \fi
}
\def\@p@@sfile#1{\@p@@sfigure{#1}}
\def\@p@@sbbllx#1{
		\@bbllxtrue
		\dimen100=#1
		\edef\@p@sbbllx{\number\dimen100}
}
\def\@p@@sbblly#1{
		\@bbllytrue
		\dimen100=#1
		\edef\@p@sbblly{\number\dimen100}
}
\def\@p@@sbburx#1{
		\@bburxtrue
		\dimen100=#1
		\edef\@p@sbburx{\number\dimen100}
}
\def\@p@@sbbury#1{
		\@bburytrue
		\dimen100=#1
		\edef\@p@sbbury{\number\dimen100}
}
\def\@p@@sheight#1{
		\@heighttrue
		\dimen100=#1
   		\edef\@p@sheight{\number\dimen100}
}
\def\@p@@swidth#1{
		\@widthtrue
		\dimen100=#1
		\edef\@p@swidth{\number\dimen100}
}
\def\@p@@srheight#1{
		\@rheighttrue
		\dimen100=#1
		\edef\@p@srheight{\number\dimen100}
}
\def\@p@@srwidth#1{
		\@rwidthtrue
		\dimen100=#1
		\edef\@p@srwidth{\number\dimen100}
}
\def\@p@@sangle#1{
		\@angletrue
		\edef\@p@sangle{#1} 
}
\def\@p@@ssilent#1{ 
		\@verbosefalse
}
\def\@p@@sprolog#1{\@prologfiletrue\def\@prologfileval{#1}}
\def\@p@@spostlog#1{\@postlogfiletrue\def\@postlogfileval{#1}}
\def\@cs@name#1{\csname #1\endcsname}
\def\@setparms#1=#2,{\@cs@name{@p@@s#1}{#2}}
\def\ps@init@parms{
		\@bbllxfalse \@bbllyfalse
		\@bburxfalse \@bburyfalse
		\@heightfalse \@widthfalse
		\@rheightfalse \@rwidthfalse
		\def\@p@sbbllx{}\def\@p@sbblly{}
		\def\@p@sbburx{}\def\@p@sbbury{}
		\def\@p@sheight{}\def\@p@swidth{}
		\def\@p@srheight{}\def\@p@srwidth{}
		\def\@p@sangle{0}
		\def\@p@sfile{} \def\@p@sbbfile{}
		\def\@p@scost{10}
		\def\@sc{}
		\@prologfilefalse
		\@postlogfilefalse
		\@clipfalse
		\if@noisy
			\@verbosetrue
		\else
			\@verbosefalse
		\fi
}
\def\parse@ps@parms#1{
	 	\@psdo\@psfiga:=#1\do
		   {\expandafter\@setparms\@psfiga,}}
\newif\ifno@bb
\def\bb@missing{
	\if@verbose{
		\ps@typeout{psfig: searching \@p@sbbfile \space  for bounding box}
	}\fi
	\no@bbtrue
	\epsf@getbb{\@p@sbbfile}
        \ifno@bb \else \bb@cull\epsf@llx\epsf@lly\epsf@urx\epsf@ury\fi
}	
\def\bb@cull#1#2#3#4{
	\dimen100=#1 bp\edef\@p@sbbllx{\number\dimen100}
	\dimen100=#2 bp\edef\@p@sbblly{\number\dimen100}
	\dimen100=#3 bp\edef\@p@sbburx{\number\dimen100}
	\dimen100=#4 bp\edef\@p@sbbury{\number\dimen100}
	\no@bbfalse
}
\newdimen\p@intvaluex
\newdimen\p@intvaluey
\def\rotate@#1#2{{\dimen0=#1 sp\dimen1=#2 sp
		  \global\p@intvaluex=\cosine\dimen0
		  \dimen3=\sine\dimen1
		  \global\advance\p@intvaluex by -\dimen3
		  \global\p@intvaluey=\sine\dimen0
		  \dimen3=\cosine\dimen1
		  \global\advance\p@intvaluey by \dimen3
		  }}
\def\compute@bb{
		\no@bbfalse
		\if@bbllx \else \no@bbtrue \fi
		\if@bblly \else \no@bbtrue \fi
		\if@bburx \else \no@bbtrue \fi
		\if@bbury \else \no@bbtrue \fi
		\ifno@bb \bb@missing \fi
		\ifno@bb \ps@typeout{FATAL ERROR: no bb supplied or found}
			\no-bb-error
		\fi
		\count203=\@p@sbburx
		\count204=\@p@sbbury
		\advance\count203 by -\@p@sbbllx
		\advance\count204 by -\@p@sbblly
		\edef\ps@bbw{\number\count203}
		\edef\ps@bbh{\number\count204}
		\if@angle 
			\Sine{\@p@sangle}\Cosine{\@p@sangle}
	        	{\dimen100=\maxdimen\xdef\r@p@sbbllx{\number\dimen100}
					    \xdef\r@p@sbblly{\number\dimen100}
			                    \xdef\r@p@sbburx{-\number\dimen100}
					    \xdef\r@p@sbbury{-\number\dimen100}}
                        \def\minmaxtest{
			   \ifnum\number\p@intvaluex<\r@p@sbbllx
			      \xdef\r@p@sbbllx{\number\p@intvaluex}\fi
			   \ifnum\number\p@intvaluex>\r@p@sbburx
			      \xdef\r@p@sbburx{\number\p@intvaluex}\fi
			   \ifnum\number\p@intvaluey<\r@p@sbblly
			      \xdef\r@p@sbblly{\number\p@intvaluey}\fi
			   \ifnum\number\p@intvaluey>\r@p@sbbury
			      \xdef\r@p@sbbury{\number\p@intvaluey}\fi
			   }
			\rotate@{\@p@sbbllx}{\@p@sbblly}
			\minmaxtest
			\rotate@{\@p@sbbllx}{\@p@sbbury}
			\minmaxtest
			\rotate@{\@p@sbburx}{\@p@sbblly}
			\minmaxtest
			\rotate@{\@p@sbburx}{\@p@sbbury}
			\minmaxtest
			\edef\@p@sbbllx{\r@p@sbbllx}\edef\@p@sbblly{\r@p@sbblly}
			\edef\@p@sbburx{\r@p@sbburx}\edef\@p@sbbury{\r@p@sbbury}
		\fi
		\count203=\@p@sbburx
		\count204=\@p@sbbury
		\advance\count203 by -\@p@sbbllx
		\advance\count204 by -\@p@sbblly
		\edef\@bbw{\number\count203}
		\edef\@bbh{\number\count204}
}
\def\in@hundreds#1#2#3{\count240=#2 \count241=#3
		     \count100=\count240	
		     \divide\count100 by \count241
		     \count101=\count100
		     \multiply\count101 by \count241
		     \advance\count240 by -\count101
		     \multiply\count240 by 10
		     \count101=\count240	
		     \divide\count101 by \count241
		     \count102=\count101
		     \multiply\count102 by \count241
		     \advance\count240 by -\count102
		     \multiply\count240 by 10
		     \count102=\count240	
		     \divide\count102 by \count241
		     \count200=#1\count205=0
		     \count201=\count200
			\multiply\count201 by \count100
		 	\advance\count205 by \count201
		     \count201=\count200
			\divide\count201 by 10
			\multiply\count201 by \count101
			\advance\count205 by \count201
		     \count201=\count200
			\divide\count201 by 100
			\multiply\count201 by \count102
			\advance\count205 by \count201
		     \edef\@result{\number\count205}
}
\def\compute@wfromh{
		\in@hundreds{\@p@sheight}{\@bbw}{\@bbh}
		\edef\@p@swidth{\@result}
}
\def\compute@hfromw{
	        \in@hundreds{\@p@swidth}{\@bbh}{\@bbw}
		\edef\@p@sheight{\@result}
}
\def\compute@handw{
		\if@height 
			\if@width
			\else
				\compute@wfromh
			\fi
		\else 
			\if@width
				\compute@hfromw
			\else
				\edef\@p@sheight{\@bbh}
				\edef\@p@swidth{\@bbw}
			\fi
		\fi
}
\def\compute@resv{
		\if@rheight \else \edef\@p@srheight{\@p@sheight} \fi
		\if@rwidth \else \edef\@p@srwidth{\@p@swidth} \fi
}
\def\compute@sizes{
	\compute@bb
	\if@scalefirst\if@angle
	\if@width
	   \in@hundreds{\@p@swidth}{\@bbw}{\ps@bbw}
	   \edef\@p@swidth{\@result}
	\fi
	\if@height
	   \in@hundreds{\@p@sheight}{\@bbh}{\ps@bbh}
	   \edef\@p@sheight{\@result}
	\fi
	\fi\fi
	\compute@handw
	\compute@resv}
\def\OzTeXSpecials{
	\special{empty.ps /@isp {true} def}
	\special{empty.ps \@p@swidth \space \@p@sheight \space
			\@p@sbbllx \space \@p@sbblly \space
			\@p@sbburx \space \@p@sbbury \space
			startTexFig \space }
	\if@clip{
		\if@verbose{
			\ps@typeout{(clip)}
		}\fi
		\special{empty.ps doclip \space }
	}\fi
	\if@angle{
		\if@verbose{
			\ps@typeout{(rotate)}
		}\fi
		\special {empty.ps \@p@sangle \space rotate \space} 
	}\fi
	\if@prologfile
	    \special{\@prologfileval \space } \fi
	\if@decmpr{
		\if@verbose{
			\ps@typeout{psfig: Compression not available
			in OzTeX version \space }
		}\fi
	}\else{
		\if@verbose{
			\ps@typeout{psfig: including \@p@sfile \space }
		}\fi
		\special{epsf=\ps@predir\@p@sfile \space }
	}\fi
	\if@postlogfile
	    \special{\@postlogfileval \space } \fi
	\special{empty.ps /@isp {false} def}
}
\def\DvipsSpecials{
	\special{ps::[begin] 	\@p@swidth \space \@p@sheight \space
			\@p@sbbllx \space \@p@sbblly \space
			\@p@sbburx \space \@p@sbbury \space
			startTexFig \space }
	\if@clip{
		\if@verbose{
			\ps@typeout{(clip)}
		}\fi
		\special{ps:: doclip \space }
	}\fi
	\if@angle
		\if@verbose{
			\ps@typeout{(clip)}
		}\fi
		\special {ps:: \@p@sangle \space rotate \space} 
	\fi
	\if@prologfile
	    \special{ps: plotfile \@prologfileval \space } \fi
	\if@decmpr{
		\if@verbose{
			\ps@typeout{psfig: including \@p@sfile.Z \space }
		}\fi
		\special{ps: plotfile "`zcat \@p@sfile.Z" \space }
	}\else{
		\if@verbose{
			\ps@typeout{psfig: including \@p@sfile \space }
		}\fi
		\special{ps: plotfile \@p@sfile \space }
	}\fi
	\if@postlogfile
	    \special{ps: plotfile \@postlogfileval \space } \fi
	\special{ps::[end] endTexFig \space }
}
\def\psfig#1{\vbox {
	\ps@init@parms
	\parse@ps@parms{#1}
	\compute@sizes
	\ifnum\@p@scost<\@psdraft{
		\PsfigSpecials 
		\vbox to \@p@srheight sp{
			\hbox to \@p@srwidth sp{
				\hss
			}
		\vss
		}
	}\else{
		\if@draftbox{		
			\hbox{\fbox{\vbox to \@p@srheight sp{
			\vss
			\hbox to \@p@srwidth sp{ \hss 
			 \hss }
			\vss
			}}}
		}\else{
			\vbox to \@p@srheight sp{
			\vss
			\hbox to \@p@srwidth sp{\hss}
			\vss
			}
		}\fi

	}\fi
}}
\psfigRestoreAt
\setDriver
\let\@=\LaTeXAtSign